\documentclass[10pt, conference, letterpaper]{IEEEtran}

\usepackage{amsmath}

\usepackage{cite}
\usepackage {color}
\usepackage{siunitx}
\usepackage{textcomp}
\usepackage{multirow}
\usepackage{graphicx}
\usepackage{subfigure}
\usepackage{float}
\usepackage{url}
\usepackage{booktabs}
\usepackage{hyperref}
\hypersetup{
    hidelinks,
    linkcolor=black,
    urlcolor=magenta,
    citecolor=black,
}
\usepackage{siunitx}
\usepackage{bbding}
\usepackage{gensymb}
\usepackage{pifont}
\usepackage{arydshln}
\usepackage{diagbox}
\usepackage{gensymb}
\usepackage{fontawesome}
\usepackage{enumitem}
\usepackage{xspace}
\usepackage{sansmath}
\usepackage{makecell}
\usepackage[justification=centering]{caption}
\usepackage[ruled,linesnumbered]{algorithm2e}
\usepackage{tikz}
\usepackage{fp}
\usepackage{algpseudocode}
\usepackage{ragged2e}
\usepackage{mathrsfs}
\usepackage{amsthm}
\usepackage{multirow}
\usepackage{colortbl}
\usepackage{wasysym}
\usepackage{fontawesome}
\usepackage{xcolor}
\usepackage{threeparttable}
\newcommand{\huanqi}[1]{\textcolor[rgb]{0, 0, 0}{#1}}

\newcommand{\mingda}[1]{\textcolor[rgb]{0, 0, 0}{#1}}

\newcommand{\SystemName}{\texttt{iRadar}\xspace}

\newlength\maxlentime
\newcommand\timebar[3][orange!40]{%
  \FPeval\result{round((#3-1.5)/#2:4)}%
  \rlap{\textcolor{#1}{\hspace*{\dimexpr-\tabcolsep+.5\arrayrulewidth}%
        \rule[-.05\ht\strutbox]{\result\maxlentime}{.95\ht\strutbox}}}%
  \makebox[\dimexpr\maxlentime-0.2\tabcolsep+\arrayrulewidth][r]{#3}}

\def\headertime{New}
\definecolor{headerColor}{RGB}{173, 216, 230}
\definecolor{rowColor1}{RGB}{245, 245, 245}
\definecolor{rowColor2}{RGB}{224, 224, 224}
\begin{document}

\title{iRadar: Synthesizing Millimeter-Waves from Wearable Inertial Inputs for Human Gesture Sensing}
\DeclareRobustCommand*{\IEEEauthorrefmark}[1]{%
  \raisebox{0pt}[0pt][0pt]{\textsuperscript{\footnotesize #1}}%
}
\author{\IEEEauthorblockN{Huanqi Yang\IEEEauthorrefmark{1},
Mingda Han\IEEEauthorrefmark{2},
Xinyue Li\IEEEauthorrefmark{3},
Di Duan\IEEEauthorrefmark{1},
Tianxing Li\IEEEauthorrefmark{4}, 
Weitao Xu\thanks{* Weitao Xu is the corresponding author.}\IEEEauthorrefmark{1,*}
}
\IEEEauthorblockA{
\IEEEauthorrefmark{1}City University of Hong Kong, \IEEEauthorrefmark{2}Shandong University,
}
\IEEEauthorblockA{
 \IEEEauthorrefmark{3}Xidian University, \IEEEauthorrefmark{4}Michigan State University
}
\vspace{-0.3in}
}
\maketitle

\begin{abstract}
Millimeter-wave (mmWave) radar-based gesture recognition is gaining attention as a key technology to enable intuitive human-machine interaction. Nevertheless, the significant challenge lies in obtaining large-scale, high-quality mmWave gesture datasets.
To tackle this problem, we present \SystemName, a novel cross-modal gesture recognition framework that employs Inertial Measurement Unit (IMU) data to synthesize the radar signals generated by the corresponding gestures.
The key idea is to exploit the IMU signals, \huanqi{which are commonly available in contemporary wearable devices}, to synthesize the radar signals that would be produced if the same gesture was performed in front of a mmWave radar. However, several technical obstacles must be overcome due to the differences between mmWave and IMU signals, the noisy gesture sensing of mmWave radar, and the dynamics of human gestures.
Firstly, we develop a method for processing IMU and mmWave data that can consistently extract critical gesture features. 
Secondly, we propose a diffusion-based IMU-to-radar translation model that accurately transforms IMU data into mmWave data. Lastly, we devise a novel transformer model to enhance gesture recognition performance.
We thoroughly evaluate \SystemName, involving 18 gestures and 30 subjects in three scenarios, using five wearable devices. Experimental results demonstrate that \SystemName consistently achieves 99.82\% Top-3 accuracy across diverse scenarios.  

\end{abstract}


\section{Introduction}
\label{sec:intro}
\subsection{Background and Motivation}
Radio Frequency (RF)--based gesture recognition has attracted significant attention due to its ability to enable contactless and device-free human-machine interaction. \huanqi{A prime example is the utilization of millimeter-wave (mmWave) signals from frequency-modulated continuous-wave (FMCW) radar for gesture recognition. This exploits that each gesture has a unique pattern, and mmWave signals can capture these differences.} The applications of mmWave-based gesture recognition extend to diverse fields such as smart homes, autonomous driving, and interactive gaming~\cite{liu2020real,smith2018gesture,akbar2023cross}. 

Despite its potential, mmWave radar-based gesture recognition, like many other RF-based sensing tasks, confronts a fundamental challenge that requires extensive training with prior instances of individuals performing gestures in the same settings~\cite{abdelnasser2015wigest,he2015wig,palipana2021pantomime}. This requirement poses difficulties for the practical deployment of this technology in real-world scenarios. For instance, in interactive gaming scenarios, a radar-based gesture recognition system may struggle to accurately identify gestures that have not been previously recorded by mmWave radar.
To address this issue, recent studies have explored the use of transfer learning~\cite{liu2022mtranssee} and domain adaptation~\cite{bhalla2021imu2doppler, li2022towards} to reduce the training burden, primarily focusing on minimizing the required training instances. While these approaches have shown promising results, it is crucial to acknowledge that they still necessitate prior radar data collection, which carries two key limitations: 1) the deployment of radar devices in the data collection area and 2) the pre-collection of gesture instances.

\begin{figure}[tbp!]
    \centering
    \includegraphics[width=0.43\textwidth]{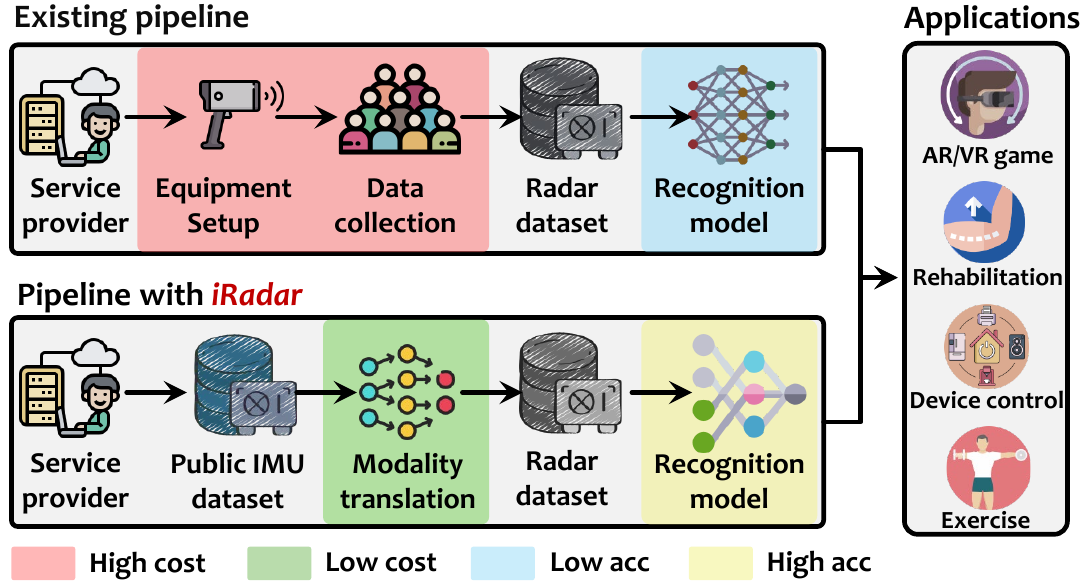}
    \caption{\textbf{Motivation of \SystemName.} }
    \label{fig:ad}
    \vspace{-0.3in}
\end{figure}

This paper introduces a novel approach that eliminates the need for prior mmWave data collection and substantially deviates from existing mmWave-based gesture recognition systems. Drawing from the recent success of diffusion models (e.g., Sora and GPT-4), we aim to investigate the viability of using alternative sensing modalities to eliminate the need for data collection. Traditional gesture recognition methods primarily rely on three sensing modalities: wireless signals~\cite{wang2020push,abdelnasser2015wigest,palipana2021pantomime}, cameras~\cite{garg2009vision,zhu2013vision,rautaray2015vision}, and wearable sensors~\cite{lu2014hand,jung2015wearable,wu2024xsolar}. This inspires us to leverage the advantage of one sensing modality (i.e., the ubiquity of cameras and wearable sensors) to overcome the data scarcity problem of mmWave-based gesture sensing. \huanqi{While previous studies have suggested the synthesis of mmWave signals from videos~\cite{ahuja2021vid2doppler,zhang2022synthesized,deng2023midas}, we found that wearable sensors provide multiple advantages over video in this context due to the following reasons.} Firstly, video-based methods still require the deployment of a camera in the data collection environment. Secondly, videos are prone to various practical factors, such as occlusion, lighting conditions, and viewpoint, which can introduce instability in the generated mmWave signals. 
Moreover, privacy concerns arise when video recording for gesture analysis is implemented. In contrast, the Inertial Measurement Unit (IMU) is widely equipped with wearable devices, such as smartwatches and rings. Therefore, utilizing the ubiquitously available IMU sensors essentially reduces device deployment costs. Additionally, there is a plethora of existing IMU-based gesture datasets, like mmGest~\cite{georgi2015recognizing} and UHH-IMU~\cite{jirak2023echo}, which can be directly translated to mmWave-based gesture datasets without extensive data collection. 
Instead of following the traditional development pipeline for mmWave gesture recognition systems, our system allows service providers to convert IMU datasets into costly-to-collect mmWave datasets with ease.
\mingda{These transformed mmWave datasets are essential for developing accurate mmWave-based gesture recognition models that meet end-user needs.}
\huanqi{To summarize, our research leverages the strengths of IMU-based systems to overcome data scarcity in mmWave-based human gesture sensing.}

\subsection{Challenges and Contributions}

\noindent\textbf{Challenge 1: Fundamental discrepancies between IMU and mmWave signals.} The signal properties of IMUs and mmWave radars are inherently distinct.  To begin with, IMUs detect the inertial forces and joint rotations associated with a person's gestures, whereas mmWave radar devices exploit the shadowing, diffraction, reflection, and scattering effects caused by the gestures on wireless signals~\cite{xue2023towards,qian2017widar,qian2018widar2,ding2020rf}. Additionally, IMU signals are expressed as real numbers, contrasting with the complex number representation of mmWave radar signals as shown in Fig.~\ref{fig:challenge2}.
To address this challenge, we thoroughly analyze the IMU and mmWave signals related to human gestures. We employ a theoretical model to explore their fundamental relationship. Yet, due to the dynamic nature of human gesture patterns, translating IMU data into mmWave data through direct mathematical formulations is a difficult task. 
\mingda{While current diffusion models~\cite{ho2020denoising, kawar2022denoising} demonstrate strong performance in tasks such as text and vision generation, they exhibit limitations in the realm of sensor data generation.}
To bridge this gap, we propose a novel deep diffusion framework equipped with an inertial fusion module and a translation module to efficiently transform IMU data into mmWave data.

\noindent\textbf{Challenge 2: Noisy gesture sensing in mmWave radar.}
A major challenge in leveraging commercial single-chip mmWave radar units for gesture recognition lies in the accurate extraction of fine-grained features critical for interpreting subtle movements. While these devices excel at detecting target movements, they often capture a significant amount of environmental noise. 
As depicted in Fig.~\ref{fig:challenge1} (a), the Range-Doppler map exhibits substantial interference. Therefore, the Time-Frequency map derived from the Range-Doppler map still contains unavoidable noise, as demonstrated in Fig.~\ref{fig:challenge1} (b), which can obscure the nuanced gestures we aim to identify. 
To address this challenge, we propose the MC-MWIE algorithm, a sophisticated dual-stage technique designed to enhance signal clarity and resolution for gesture recognition applications. It engages a synergistic approach combining cluster analysis with morphological processing to mitigate environmental noise and refine the resolution. The result is a set of mmWave heatmaps with improved clarity, enabling accurate feature extraction.


\begin{figure}[tbp!]
\centering
\subfigure{
\begin{minipage}[t]{0.47\linewidth}
\setcounter{figure}{1}
\centering
    \includegraphics[width=1.6in]{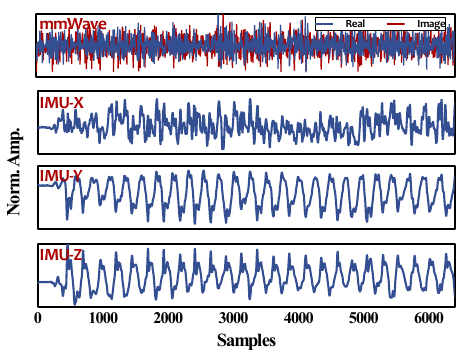}
    \caption{\textbf{Signals difference.}
    }
    \label{fig:challenge2}
\end{minipage}%
}
\subfigure{
\begin{minipage}[t]{0.47\linewidth}
\centering
    \includegraphics[width=1.6in]{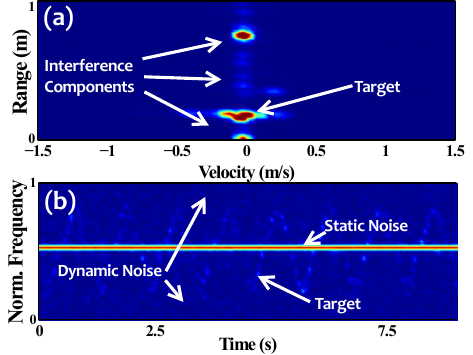}
    \caption{\textbf{\huanqi{Noisy gesture sensing.}}
    }
    \label{fig:challenge1}
\end{minipage}%
}
\vspace{-0.3in}
\end{figure}

\noindent\textbf{Challenge 3: Dynamics of human gestures.} Gestures involve the coordinated movements of multiple body parts, which encompass various actions and involve numerous joints and muscles~\cite{freivalds2011biomechanics}. Capturing the full dynamics of these complex 3D motions with a single sensing modality proves difficult. Additionally, the variability and subtlety among different gestures often impede gesture recognition accuracy.
To enhance the precision of gesture recognition, it is imperative to move beyond the traditional convolution-based method~\cite{yuan2022real,chen2022hand,liu2022mtranssee}, which are insufficient for the complexity of gesture dynamics. Transformers have demonstrated effectiveness in handling vision tasks\mingda{~\cite{dosovitskiy2020image}}; however, their direct application to mmWave heatmaps of gestures presents challenges, as the unique time-frequency properties of gesture signals differ significantly from typical image data.
In response, we introduce a novel transformer for the unique mmWave gesture heatmaps. This model boosts recognition by integrating specialized components to better capture the intricacies of gestures.

In this paper, we propose \SystemName, a \huanqi{cross-modal} \textbf{i}MU-to-\textbf{Radar} gesture recognition framework. This system eliminates the necessity for the initial deployment of mmWave radar devices and the collection of explicit data. This progress pushes mmWave-enabled gesture recognition technologies into a realm of real-world usability.
Through a comprehensive evaluation that included eighteen gestures, thirty participants, tested across three distinct settings, and utilizing five different mobile devices, \SystemName has proven its efficacy by attaining an average accuracy of 92.3\% in settings ranging from indoors and outdoors to through-obstacle scenarios. 
Our key contributions are outlined as follows:
\begin{itemize}[leftmargin=*]
\item We present \SystemName, which, to our best knowledge, is the first cross-modal IMU-to-mmWave gesture recognition framework that avoids the installation of mmWave device and explicit data collection, significantly reducing the burden of \huanqi{service providers}. 

\item \SystemName offers threefold specialized approaches, which include a \huanqi{diffusion-driven translation method, a mmWave heatmap enhancement method, and a Doppler transformer recognition method.} Collectively, these methods tackle the above challenges and ensure accurate gesture recognition.

\item We develop a system prototype and conduct extensive experiments to evaluate the performance of \SystemName across various scenarios and mobile devices. Experimental results indicate \SystemName consistently achieves an average Top-3 accuracy of 99.82\%.
\end{itemize}

\section{Preliminaries}
\label{sec:pre}

\subsection{mmWave Sensing}
\label{subsec:RFSening}
The mmWave radar utilizes the FMCW signal, often referred to as the chirp signal.
The chirp signal's frequency increases linearly with time $t$ according to the equation $f(t)=f_c+St$, where $f_c$ denotes the starting frequency and $S$ represents the frequency modulation slope~\cite{rao2017introduction,han2023mmsign}.
Assuming the amplitude of the transmitted signal at time $t$ is $A_1$, the transmitted FMCW signal $S_{{Tx}}(t)$ is expressed as
\begin{equation}\label{equ:Tx Signal}\footnotesize
    S_{_{Tx}}(t)=A_1 \cos \left[2 \pi\left(f_c t+\frac{St^2}{2}\right)\right].
\end{equation}
When the transmitted signal encounters an obstacle, such as the user's hand, at a distance $d$, the radar receives a delayed version of the transmitted signal $S_{{Tx}}(t)$, denoted as $S_{{Rx}}(t)$. This received signal can be expressed as
\begin{equation}\label{equ:Rx Signal}\footnotesize
    S_{_{Rx}}(t)=\alpha A_1 \cos \left[2 \pi\left(f_c\left(t-\tau\right)+\frac{S(t-\tau)^2}{2}\right)\right],
\end{equation}
where $\alpha $ represents the path loss, $\tau = 2d/c $ denotes the time delay, and $c$ is the speed of light.
Finally, the transmitted signal $S_{{Tx}}(t)$ is mixed with the received signal $S_{{Rx}}(t)$, and a low-pass filter is employed to extract the sum frequency components, resulting in the intermediate frequency (IF) signal:
\begin{equation}\label{equ:IF Signal}\footnotesize
    S_{_{IF}}(t)=LPF\{S_{_{Tx}}(t)\cdot S_{_{Rx}}(t)\} = A_2\cos\left(2\pi f_{_{IF}}t+\phi_{_{IF}}\right),
\end{equation}
where $A_2$ is the amplitude of the IF signal, $f_{{IF}} = 2dS/c$ is the beat frequency, and $\phi{_{IF}}$ represents the phase of the IF signal.
The FMCW mmWave radar enables the extraction of crucial target information such as range and velocity. Specifically, as depicted in Fig.~\ref{fig:fft}, the range information is determined by applying the Fast Fourier Transform (Range FFT).
The velocity information is obtained by performing the Fast Fourier Transform (Doppler FFT) on multiple IF signals spanning the slow time dimension.

\begin{figure}[tbp!]
\setcounter{figure}{3}
    \centering
    \includegraphics[width=0.45\textwidth]{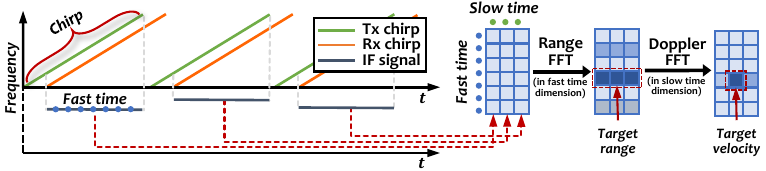}
    \caption{\textbf{FMCW signals and processing methods.} }
    \label{fig:fft}
    \vspace{-0.2in}
\end{figure}

\subsection{Human Gesture} 
\label{subsec:understandgesture}

\begin{figure}
\centering
\subfigure[Movement process.]{
\begin{minipage}[t]{0.42\linewidth}
\centering
\includegraphics[width=\linewidth]{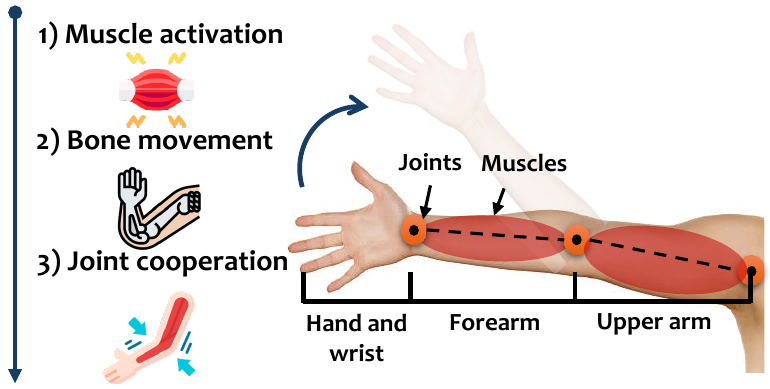}
\label{fig:understandgesture1}
\vspace{-6mm}
\end{minipage}%
\vspace{-6mm}
}%
\hspace{-6mm}
\subfigure[Skeletal muscles of arm.]{
\begin{minipage}[t]{0.46\linewidth}
\centering
\includegraphics[width=\linewidth]{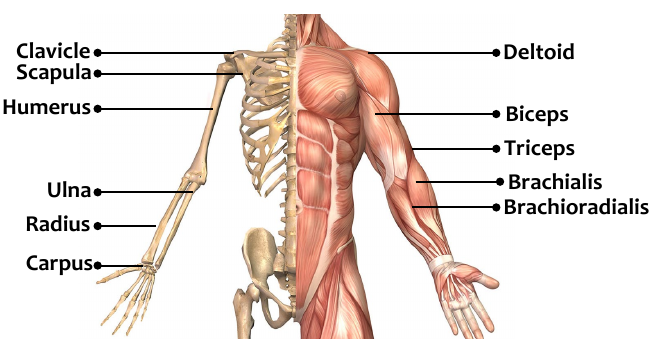}
\label{fig:understandgesture2}
\vspace{-6mm}
\end{minipage}%
\vspace{-6mm}
}%
\vspace{-1mm}
\centering
\caption{\textbf{Understanding human gesture.}}
\label{fig:understandgesture}
\vspace{-0.2in}
\end{figure}

Human gestures are characterized by the intricate movements and postures adopted by various body segments~\cite{freivalds2011biomechanics}. These gestures originate from the unique interactions among specific body components.
\huanqi{As shown in Fig.~\ref{fig:understandgesture}, the human gesture is driven by three steps.
1) Muscle activation: the biceps brachii muscle flexes the elbow, while the deltoid muscle aids in moving the arm at the shoulder joint.
2) Bone movement: the rotation of the radius over the ulna allows pronation and supination of the forearm.
3) Joint cooperation: the combined action of the shoulder's ball-and-socket joint, the elbow's hinge joint, and the wrist's complex array of plane and hinge joints.}
These synchronized elements facilitate a diverse range of nonverbal expressions.

\begin{figure}[tbp!]
    \centering
    \includegraphics[width=0.41\textwidth]{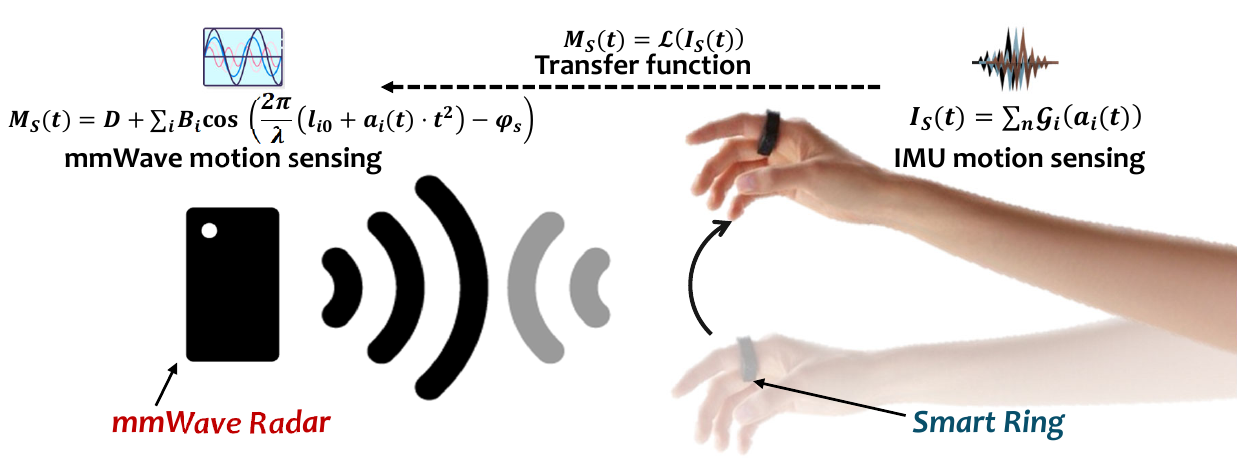}
    \caption{\textbf{Cross-modal relationship analysis.} }
    \label{fig:relationship}
    \vspace{-0.2in}
\end{figure}


\subsection{Cross-Modal Relationship Analysis}
\label{subsec:Correlation}
We now explore the relationship between mmWave and IMU gesture sensing, as shown in Fig.~\ref{fig:relationship}.

\noindent\textbf{IMU signal.} 
As detailed in Sec.~\ref{subsec:understandgesture}, human gestures involve distinct movements of specific body parts. The IMU data from an on-arm wearable device (denoted as ${I_s}(t)$, collected during these gestures) is a composite representation of the accelerations experienced by the engaged body segments: {\footnotesize${I_s}(t) = \sum_n \mathcal{G}_i(a_i(t)),$}
where $a_i(t)$ is the acceleration of the $i^{\text{th}}$ body part involved in the execution of the gesture at time $t$, and $\mathcal{G}_i(\cdot)$ represents the transfer function mapping the acceleration due to the movement of each body part to the on-arm device.

\noindent\textbf{mmWave signal.} 
As discussed in Sec.~\ref{subsec:RFSening}, the movement of human body parts during the execution of the gesture can lead to phase changes in the mmWave signal. 
Therefore, we can express the mmWave signal variations induced by different gestures mathematically as follows:
\begin{equation} \footnotesize
M(f,t) = H_0(f,t) + \sum_i A_i(f,t) e^{-j(\frac{2\pi}{\lambda} (l_{i0} + a_i(t)\cdot t^2))},
\end{equation}
where $H_0(f,t)$ represents the complex static path signal generated by the human body and surrounding environmental objects~\cite{zhang2022quantifying,zhang2022can,ji2023construct,yang2023xgait,han2024seeing,yang2023wave,cui2023mmripple,han2023mmsign}, while $A_i(f,t)$ denotes the amplitude of the dynamic path signal reflected from the $i^{\text{th}}$ arm/hand segment during gesture execution. The term $l_{i0}$ signifies the initial signal propagation distance, and $a_i(t)$ denotes the acceleration of the arm/hand segment.
Let $M_s(t)$ represent the magnitude square of the baseband signal \mingda{$M(f,t)$}. Assuming $\left|H_0(f,t)\right| \gg \left|A_i(f,t)\right|$, we can represent $M_s(t)$ as
\begin{equation}\label{eq:RF} \footnotesize
M_s(t) = D + \sum_i B_i \cos \left(\frac{2 \pi}{\lambda} (l_{i0} + a_i(t)\cdot t^2) - \varphi_s\right),
\end{equation}
where $D = \left|H_0(f,t)\right|^2 + \sum_i\left|A_i(f,t)\right|^2$ represents the DC component of $M_s(t)$, $B_i = 2\left|H_0(f,t)\cdot A_i(f,t)\right|$, and $\varphi_s$ denotes the phase of the complex static path signal. 


\noindent\textbf{Relationship analysis.}
As mentioned above,
since both ${I_s}(t)$ and $M_s(t)$ are functions of acceleration $a_i(t)$,
${I_s}(t)$ incorporates frequency components that either coincide with or closely resemble the frequencies present in $R(t)$.
Both data types are defined by functions that use acceleration as the independent variable, suggesting the potential to transform IMU data into mmWave data through a non-linear function $\mathcal{L}(\cdot)$: {\footnotesize$M_s(t) = \mathcal{L}({I_s}(t)).$}
\begin{figure}[tbp!]
    \centering
    \includegraphics[width=0.41\textwidth]{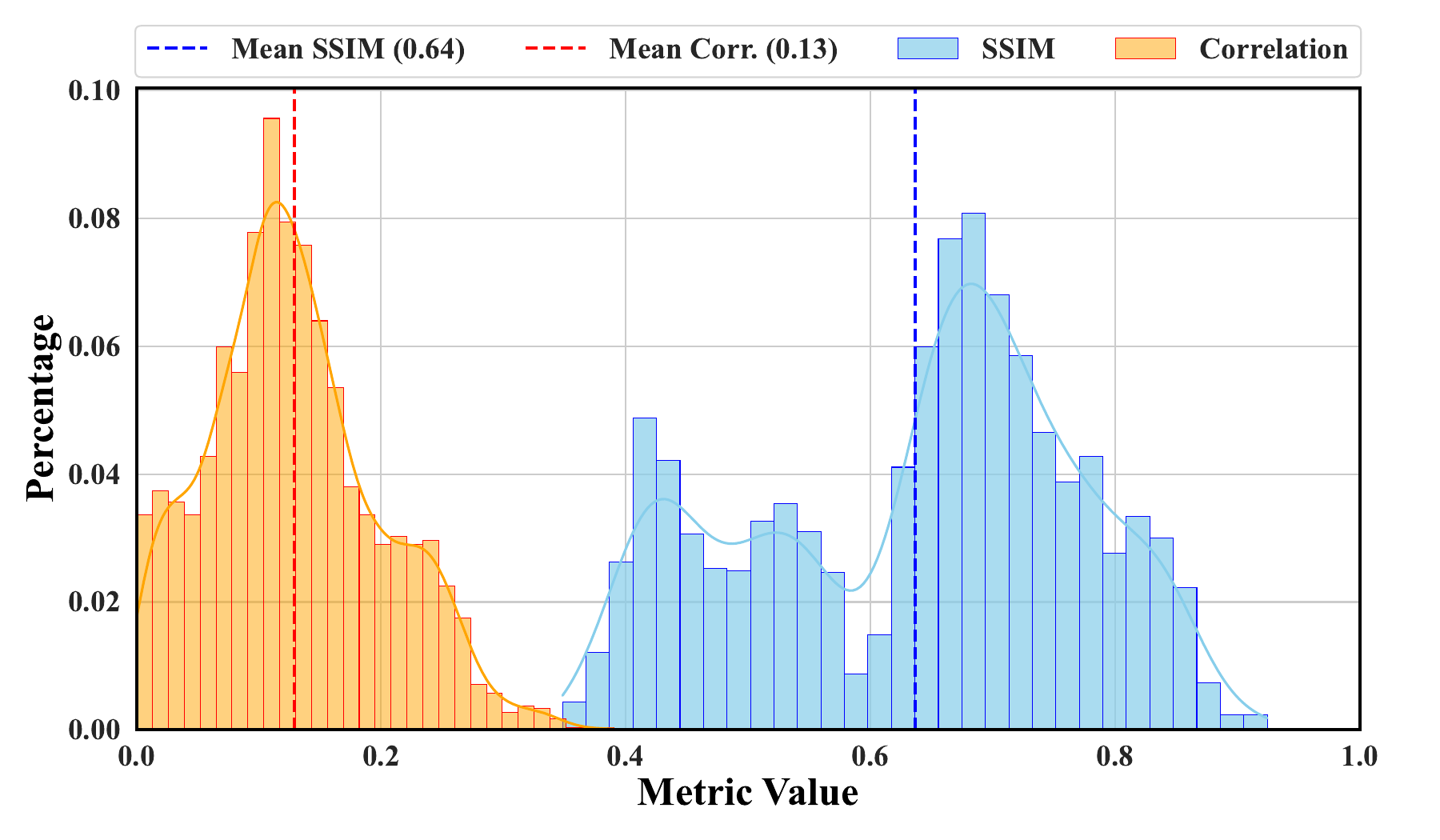}
    \vspace{-0.1in}
    \caption{\textbf{Similarity distribution between IMU and mmWave features.}} 
    \vspace{-0.3in}
    \label{fig:ssim_dis}
\end{figure}
We first examined the linear relationship between the features extracted from IMU spectrograms and mmWave heatmaps by calculating the coefficient of \huanqi{linear correlation}. This analysis revealed a mean correlation value of 0.13, as illustrated by the orange line in Fig.~\ref{fig:ssim_dis}. This low correlation coefficient suggests a weak linear relationship between these modalities. To further investigate the structural similarities between them, we utilized the Structural Similarity Index Measure (SSIM)~\cite{wang2004image}, which measures the similarity between two features based on an understanding of visual perception. In our analysis, the mean SSIM between IMU and mmWave features is 0.64, as shown in the blue line in Fig.~\ref{fig:ssim_dis}, indicating a moderate structural similarity. 
The above results indicate the complexity of defining $\mathcal{L}(\cdot)$ that accurately characterizes the relationship between the IMU spectrogram and mmWave heatmap features. Due to this complexity and the inherent non-linearity between the two modalities, along with other complicating factors such as noise, signal attenuation, and multi-path effects~\cite{zhang2014accelerometer,adib2013see}, conventional mathematical approaches to defining $\mathcal{L}(\cdot)$ are insufficient. 
Therefore, we propose employing diffusion techniques, which are well-suited for capturing complex, non-linear relationships to establish an accurate mapping between them.

\section{System Design}

Fig.~\ref{fig:system} shows \SystemName's overview, comprising three layers.

\noindent\textbf{Cross-Modal Learning Layer.}
In the cross-modal learning layer, we devise a deep diffusion model aimed at converting the IMU feature into the mmWave feature. Initially, the raw IMU data and mmWave data undergo preprocessing to mitigate noise. Subsequently, IMU and mmWave features are generated from each dataset using our proposed signal processing algorithms. Finally, the proposed diffusion model is trained for IMU-to-mmWave translation.

\begin{figure}[tbp!]
    \centering
    \includegraphics[width=0.41\textwidth]{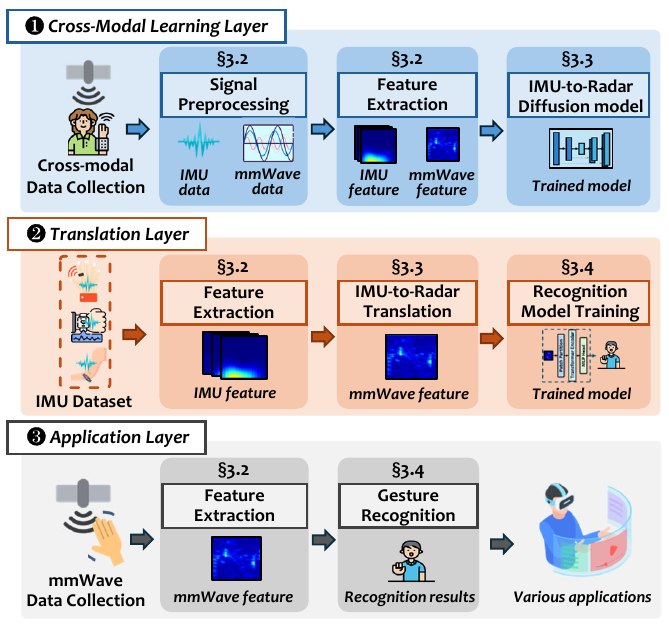}
    \caption{\textbf{\SystemName overview.} }
    \label{fig:system}
    \vspace{-0.2in}
\end{figure}
\noindent\textbf{Translation Layer.}
In the translation layer, service providers have the option to utilize either publicly available or proprietary IMU data. This data is then translated into mmWave heatmaps using the trained IMU-to-Radar diffusion model. Once the translation is complete, the resulting mmWave heatmaps are used to train the gesture recognition network.

\noindent\textbf{Application Layer.}
In the application layer, users can conveniently utilize mmWave radar-embedded smart devices to directly capture mmWave gesture data and conduct real-time gesture recognition for diverse applications. 


\subsection{Feature Extraction}
\subsubsection{mmWave Heatmap Generation}
As illustrated in Sec.~\ref{subsec:RFSening}, the IF signal obtained by mixing is used for sensing gesture. 
In addition to the user's gesture information, the IF signal contains a lot of static noise generated by static objects such as walls, tables, and chairs as shown in Fig.~\ref{fig:rawRDM}. For each frame, we use the average of all IF signals as the static noise vector and subtract this static noise vector from each IF signal to obtain the denoised data as shown in Fig.~\ref{fig:DeRDM}. 
Then, we use the Range FFT and Doppler FFT to obtain a Range-Doppler Map (RDM) for each frame, which reflects the range and velocity information of the user while performing gesture in the current frame. Finally, to obtain the velocity change information during the execution of the gesture, we transform the RDMs of all frames into a 2D time-velocity feature map.
As discussed in Sec.~\ref{subsec:RFSening}, the Doppler FFT responds to changes in phase difference, which are proportional to frequency. Therefore, we can derive the normalized frequencies by normalizing the extracted velocities \mingda{as shown in Fig.~\ref{fig:modwt_result}(a)}.

\begin{figure}[tbp!]
\centering
\subfigure{
\begin{minipage}[t]{0.47\linewidth}
\setcounter{figure}{8}
\centering
    \includegraphics[width=1.5in]{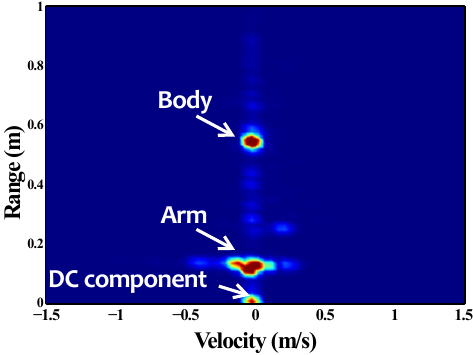}
    \caption{\textbf{Raw RDM.}
    }
    \label{fig:rawRDM}
\end{minipage}%
}
\subfigure{
\begin{minipage}[t]{0.47\linewidth}
\centering
    \includegraphics[width=1.5in]{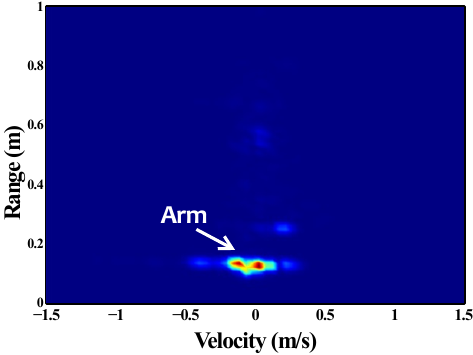}
    \caption{\textbf{Denoised RDM.}
    }
    \label{fig:DeRDM}
\end{minipage}%
}
\vspace{-0.2in}
\end{figure}

\begin{figure}[tbp!]
\centering
\subfigure{
\begin{minipage}[t]{0.47\linewidth}
\centering\setcounter{figure}{10}
    \includegraphics[width=1.5in]{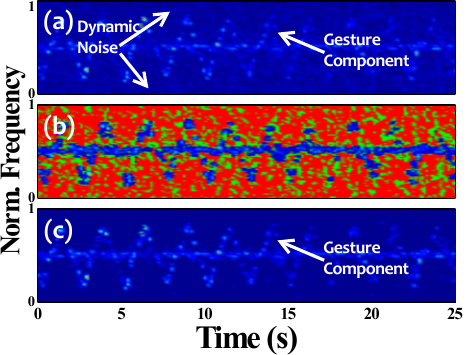}
    \caption{\textbf{MC-MHWE process.}
    }
    \label{fig:modwt_result}
\end{minipage}%
}
\subfigure{
\begin{minipage}[t]{0.47\linewidth}

\centering
    \includegraphics[width=1.5in]{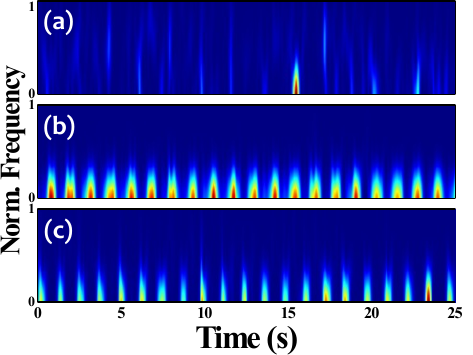}
    \caption{\textbf{Extracted IMU feature.}
    }
    \label{fig:modwt_imu}
\end{minipage}%
}
\vspace{-0.3in}
\end{figure}


\begin{figure*}[tbp!]
    \centering
    \setcounter{figure}{12}
    \includegraphics[width=1\textwidth]{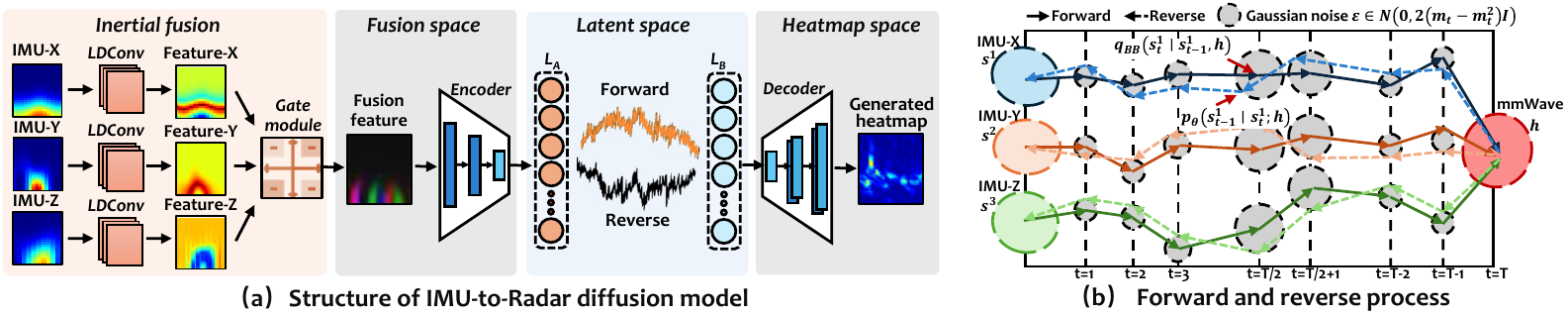}
    \caption{\textbf{I2R diffusion model.}} 
    \vspace{-0.3in}
    \label{fig:I2R}
\end{figure*}

The resulting time-velocity feature map may exhibit some noisy components stemming from the superposition effects of RDMs, which can obscure significant velocity changes in the gestures. To address this issue, we propose Morphological Clustering for mmWave Heatmap Enhancement (MC-MWHE), a mmWave feature enhancement algorithm based on image morphological operations.
Specifically, we first apply Gaussian blurring to the original feature map to mitigate noise. Subsequently, we utilize K-means clustering to segregate the pixels into two discrete categories, as depicted in Fig.~\ref{fig:modwt_result}(b), where blue represents the gesture component and red indicates the dynamic noise component. Subsequently, pixels are reassigned based on the predominant clusters to accentuate the primary features. Following this, the feature map undergoes conversion to grayscale and binarization using a mean threshold. Finally, morphological closure operations are employed to bridge the discontinuities in the gesture component, which are marked in green. The final enhanced mmWave time-velocity heatmap is illustrated in Fig.~\ref{fig:modwt_result}(c).

\subsubsection{IMU Spectrogram Generation}
\label{subsec:IMU_spe_gen}
In \SystemName, to isolate gesture-related signals, the acceleration data along the X, Y, and Z axes are decomposed into four levels utilizing maximal overlap discrete wavelet transform (MODWT).
Finally, the three processed IMU signals are employed to derive three spectrograms using short-time Fourier transform (STFT), which are shown in Fig.~\ref{fig:modwt_imu}(a)-(c), respectively. 
The obtained spectrograms from the three directions are utilized as the inputs of the inertial fusion module in Sec.~\ref{subsec:speFusion}.

\subsection{IMU-to-Radar Diffusion Model}
\label{sec:spec2spec}
\huanqi{This section introduces IMU-to-Radar (I2R)}, a diffusion model designed to convert IMU spectrograms to mmWave heatmaps. Fig.~\ref{fig:I2R} shows the I2R structure, which consists of an inertial fusion module and a translation module.

\subsubsection{IMU Inertial Fusion}
\label{subsec:speFusion}
IMU data contains a wealth of information related to signal frequency and motion intensity. Furthermore, the frequency spectrum shapes of IMU data can differ significantly between gestures. As a result, effectively modeling and extracting features from the IMU spectrogram poses a significant challenge. Consequently, we propose an inertial fusion module comprising a Learnable Dilated Convolutional Neural Network (LDCNN)—a novel convolutional approach to extract the features from IMU spectrograms. Additionally, a gating mechanism assigns varying weights to the features extracted by the LDCNN.


\noindent\textbf{LDCNN-based feature extraction.} Recall that the standard convolution operation
which is characterized as {\footnotesize$ O(\sigma) = \sum_{\sigma' \in S} I(\sigma + \sigma') * K(\sigma'),$}
where $O(\sigma)$ is the output feature map, $I(\sigma)$ is the input feature map, $K(\sigma')$ is the convolution kernel, and $S$ is the neighborhood around the pixel $\sigma$.
DCNN (Dilated Convolutional Neural Network) is designed to expand the convolutional kernel by periodically inserting spaces (i.e., zeros) between the kernel elements~\cite{yu2015multi}.
As a result, the spacing between the elements' dilation rate can be described as {\footnotesize $O(\sigma) = \sum_{\sigma' \in S} I(\sigma + d \cdot \sigma') * K(\sigma'),$}
\huanqi{where $I$ represents the input feature map, $K$ is the dilated convolution kernel, and $d$ is the dilation rate.}
In LDCNN, the positions of non-zero elements within the convolutional kernel are learned using a gradient-based method. However, since the positions in the kernel are integer values, it poses a challenge in terms of differentiability. To overcome this issue, we utilize interpolation.
The main motivation behind LDCNN is to explore the potential of enhancing the fixed grid imposed by the standard DConv by learning the spacing in an input-independent manner. Unlike the grid-like arrangement of convolutional kernel elements in standard and dilated convolutions, LDCNN allows for a flexible number of kernel elements~\cite{khalfaoui2023dilated}: {\footnotesize$ O(\sigma) = \sum_{\sigma' \in S} I(\sigma + L(\sigma) \cdot \sigma') * K(\sigma'),$}
\huanqi{where $L(\sigma)$ is the learnable dilation rate function, which is updated through backpropagation.}

\noindent\textbf{Gate module.} 
Following the LDCNN stage, the extracted features are fed into a gate module. The purpose of this module is to selectively fuse information, leveraging a gating mechanism to filter and combine pertinent features from the three orthogonal axes of IMU data. For IMU spectrogram features refined by the LDCNN, the gate module operates as {\footnotesize$G(s^1, s^2, s^3) = F(LDCNN(s^1, s^2, s^3); \eta),$}
where $s^1, s^2, s^3$ is the input IMU spectrograms, $G(\cdot)$ denotes the gated feature output, $F(\cdot)$ is a fully connected fusion layer that integrates the derived features, and $\eta$ is the learned parameters.

\subsubsection{Bridge Diffusion-based Translation}
Based on Brownian Bridge Diffusion (BBDM)~\cite{li2023bbdm},
our approach incorporates a bilateral framework specifically designed to bridge the gap between IMU and mmWave data. By considering both the correlation and the unique characteristics of IMU spectrograms and mmWave heatmaps, BBDM effectively captures the nuanced mapping relationship between the two. Fig.~\ref{fig:I2R} (b) illustrates the mathematical process proposed by the translation method, which includes the forward process and the reverse process.


\noindent\textbf{Forward process.} 
The forward process describes the diffusion of IMU spectrograms to mmWave heatmaps, which initiates from the IMU spectrograms and progressively incorporates noise and drift, gradually transitioning towards the mmWave heatmaps. The IMU spectrograms are represented by a set of inputs $s = \{s^1, s^2, s^3\}$. We let $(s, h)$ denote the paired training data from IMU spectrograms and mmWave heatmaps. We take the ground truth mmWave heatmap conditional input $h$ as its destination. It is assumed that $s$ and $h$ are approximately independent and normally distributed as $s, h \sim \mathcal{N}(\mathbf{0}, \boldsymbol{I})$. Given initial state $s_0$ (as the blue, orange, green circles illustrated in Fig.~\ref{fig:I2R} (b)), the intermediate state $s_t$ (as the grey circle illustrated in Fig.~\ref{fig:I2R} (b)) and destination state $h$ (as the red circle illustrated in Fig.~\ref{fig:I2R} (b)), the forward diffusion process of Brownian Bridge can be defined as:
\begin{equation}\footnotesize
\begin{gathered}
q_{B B}\left(s_t \mid s_0, h\right)=\mathcal{N}\left(s_t ;\left(1-m_t\right) s_0+m_t h, \delta_t I\right), \quad m_t=\frac{t}{T},
\end{gathered}
\label{eq_bbdm_1}
\end{equation}
where $T$ is the total steps of the diffusion process, $\delta_t$ is the variance. For training and inference purposes, we need to deduce the forward transition probability $q_{B B}\left(x_t \mid x_{t-1}, h\right)$ (as the grey line illustrated in Fig.~\ref{fig:I2R} (b)):
\begin{equation}\footnotesize
\begin{array}{r}
q_{B B}\left(s_t \mid s_{t-1}, h\right)=\mathcal{N}\left(s_t ; \frac{1-m_t}{1-m_{t-1}} s_{t-1}\right. \\
\left.+\left(m_t-\frac{1-m_t}{1-m_{t-1}} m_{t-1}\right) h, \delta_{t \mid t-1} I\right).
\end{array}
\label{eq_bbdm_2}
\end{equation}
According to Eq.~\ref{eq_bbdm_1}, when the diffusion process reaches the destination, i.e., $t=T$, we can get that $m_T=1$. The forward diffusion process defines a fixed mapping from IMU spectrograms to mmWave heatmaps.

\noindent\textbf{Reverse process.} 
The reverse diffusion process can be utilized to infer the possible initial IMU spectrograms that could have resulted in the observed mmWave heatmaps by exploiting UNet to learn the mapping functions between IMU spectrograms and their corresponding mmWave heatmaps by minimizing the difference. It serves as the inverse of the forward diffusion process. Starting from the mmWave heatmaps, the backward diffusion process gradually eliminates the noise and drift through reverse operations, restoring the distribution towards the IMU spectrograms. Different from the existing diffusion models, the Brownian Bridge process directly starts from the conditional input by setting ${s_T}={h}$. The reverse process aims to predict $\boldsymbol{s}_{t-1}$ based on ${s_t}$ :
\begin{equation}
p_\theta\left(s_{t-1} \mid s_t, {h}\right)=\mathcal{N}\left(s_{t-1} ; \mu_\theta\left(s_t, t\right), \delta_t I\right),
\end{equation}
where $\mu_\theta\left(s_t, t\right)$ is the predicted mean value of the noise, and $\tilde{\delta}_t$ is the variance of noise at each step. The mean value ${\mu}_\theta\left({s_t}, t\right)$ is required to be learned by a neural network with parameters $\theta$ based on the maximum likelihood criterion.

\begin{figure}[tbp!]
    \centering
    \includegraphics[width=0.4\textwidth]{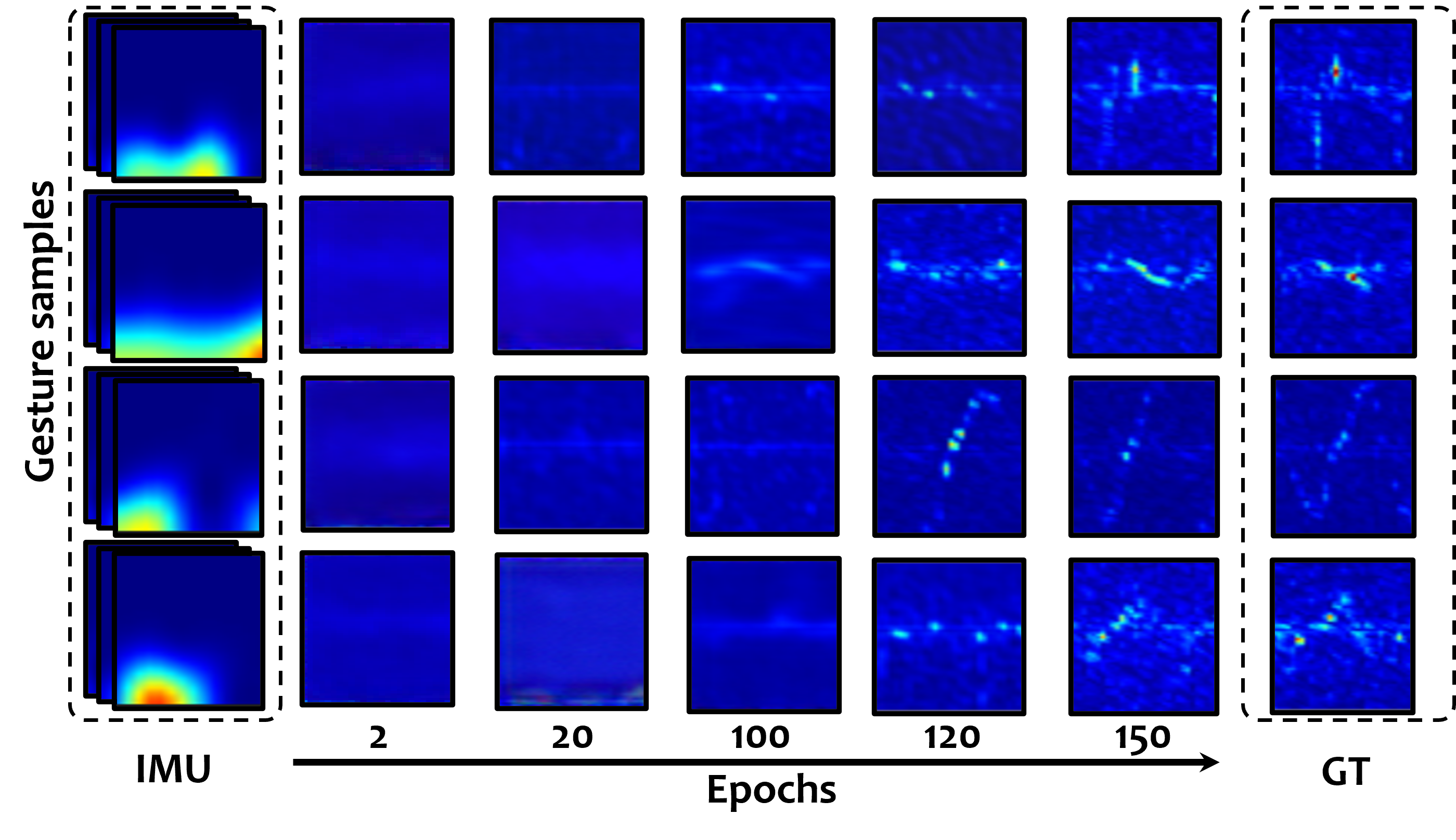}
    \caption{\textbf{Training progress for various gestures.} Sequentially from top to bottom: front raise, lateral-to-front raise, push, and forearm supination.} 
    \vspace{-0.2in}
    \label{fig:epoch}
\end{figure}

\noindent\textbf{Implementation.} 
We utilize input IMU spectrograms in the inertial fusion module as the source domain and mmWave heatmaps as the target domain to train the diffusion model. Additionally, we implement a U-Net neural network and employ it in the backward process. The U-Net architecture consists of four encoders and four decoders, utilizing ReLU as the activation function and max pooling for pooling operations. The inertial fusion module, forward process, and reverse process are closely connected as they are jointly trained to optimize the overall performance. As shown in Fig.~\ref{fig:epoch}, training progress from different gesture samples shows the convergence and stability of the model.


\subsection{Gesture Recognition}
As outlined in Sec.~\ref{sec:intro}, accurate gesture recognition is a hard task due to the inherent complexity of arm movements and the diverse array of gesture patterns. Traditional approaches to feature extraction in gesture recognition often rely on convolutional neural networks~\cite{yuan2022real,chen2022hand,liu2022mtranssee}, which may be inherently constrained by their local receptive fields and weight-sharing properties, potentially limiting their capacity to capture the global dependencies within complex gesture data.
Drawing inspiration from the success of transformer architectures~\cite{dosovitskiy2020image, lee2021vision}, we introduce a novel Doppler transformer tailored for interpreting Doppler heatmaps of gestures in the following.

\begin{figure}
    \centering
    \includegraphics[width=0.46\textwidth]{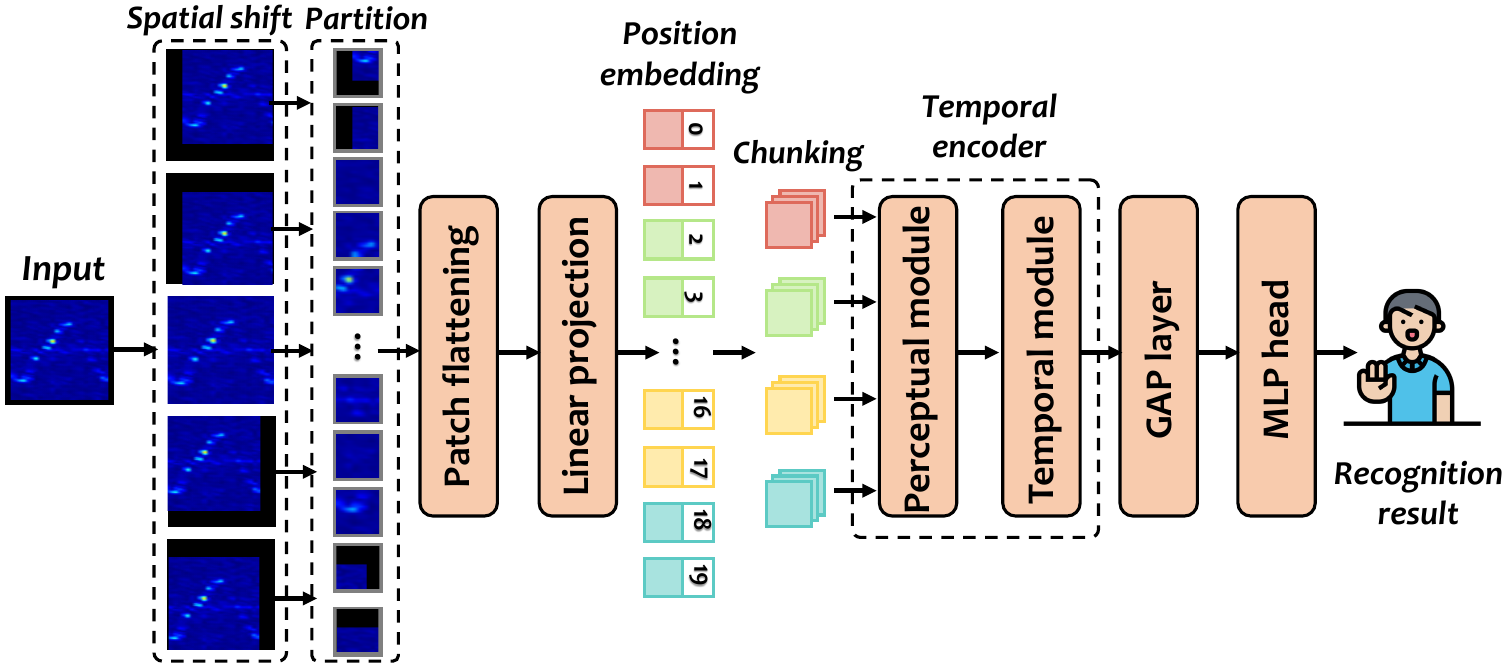}
    \caption{\textbf{Doppler transformer.}}
    \label{fig:transformer}
    \vspace{-0.3in}
\end{figure}

\noindent\textbf{Spatial heatmap shift and patch embedding.}
As shown in Fig.~\ref{fig:transformer}, our approach enhances traditional vision transformers' limited receptive fields~\cite{dosovitskiy2020image} by employing spatial heatmap patches shifted along various diagonal axes~\cite{lee2021vision}, resulting in an enriched representation of the time-doppler landscape through overlapping and merging with original heatmaps.


\noindent\textbf{Temporal attention layer.}  
To effectively learn heatmap details, we use the temporal attention mechanism to concentrate on the most significant temporal information. Specifically, we first divide the patch embedding into chunks by temporal sequence. 
Then we use a perceptual module to extract spatial features and a temporal module~\cite{didolkar2022temporal} to integrate these features over time, ensuring a comprehensive understanding of both space and time within the data.

\noindent\textbf{Implementation.} The model comprises two temporal transformer layers, each projecting patches into a 64-dimensional embedding space using a single attention head. It processes input data in chunks of eight. Training proceeds for 1000 epochs with a learning rate and weight decay both set at 0.001. 

\section{Evaluation}
\subsection{Experimental Methodology}
\begin{figure}[tbp!]
    \centering
    \includegraphics[width=0.45\textwidth]{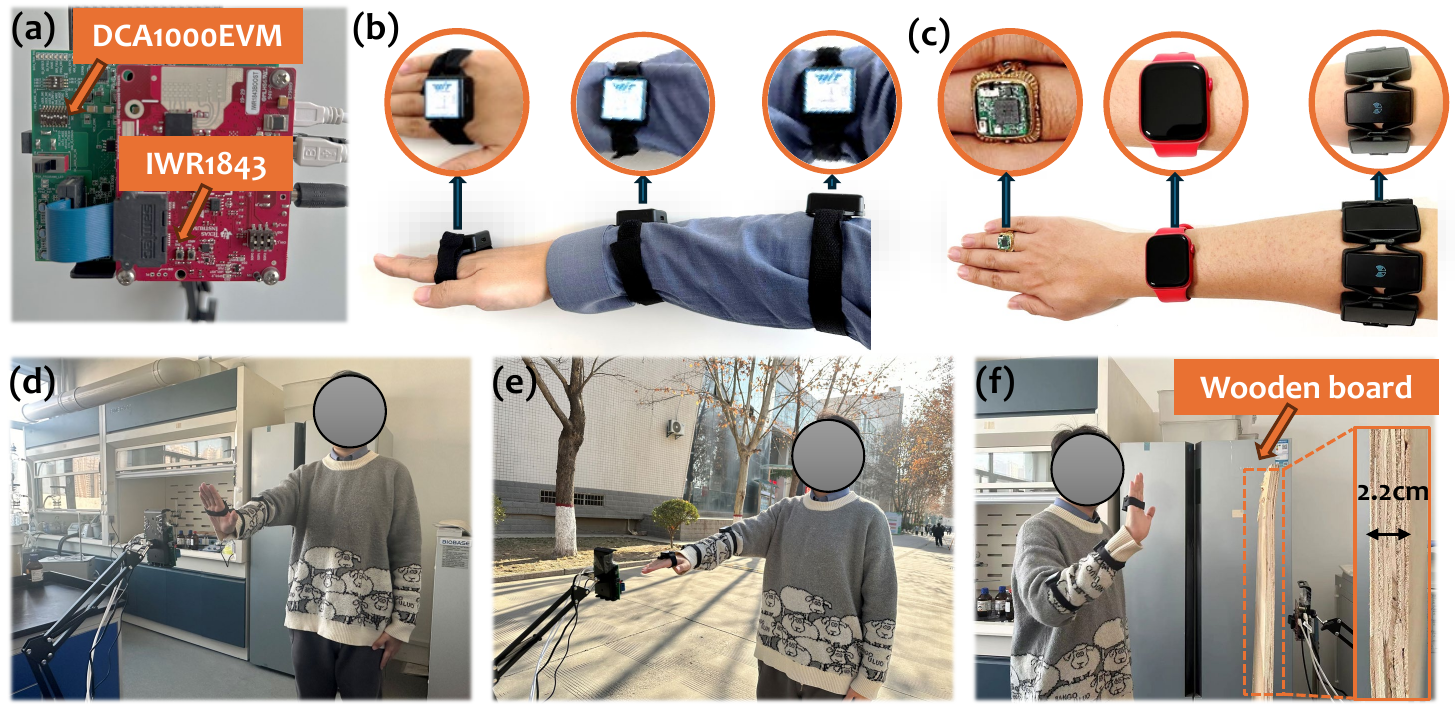}
    \caption{\textbf{Experimental setup.} }
    \label{fig:dev}
    \vspace{-0.1in}
\end{figure}

\noindent\textbf{System implementation.} The setup for evaluating \SystemName comprises the experimental devices depicted in Fig.~\ref{fig:dev}. As illustrated in Fig.~\ref{fig:dev}(a), data collection for mmWave sensing is conducted using a 77 GHz IWR1843 FMCW radar coupled with a DCA1000EVM real-time data capture adapter. Fig.~\ref{fig:dev}(b) and (c) display the WT901WIFI motion sensor and four wearable sensors employed for acquiring IMU data, essential for the training and testing of the base model. In addition, our evaluation considers a variety of wearable technologies including a smart armband, smartwatch, and smart ring. Details on these devices can be found in Tab.~\ref{tab:device}. The IMUs operate at a default sampling rate of 100 Hz.

\begin{table}[tbp!]
\caption{\textbf{Wearable device and mmWave radar specifications}}
\centering
\fontsize{6.7pt}{7.2pt}\selectfont 
\begin{threeparttable}
\setlength{\tabcolsep}{3pt}
\centering
\begin{tabular}{l|S|S|l|l}
\hline
\textbf{Name} & \textbf{CPU Freq} & \textbf{RAM} & \textbf{OS} & \textbf{IMU Model\tnote{*}} \\
\hline
WT901WIFI          & \SI{168}{\mega\byte} & N/A & N/A & IS MPU9250 \\
Smart Ring & \SI{64}{\mega\byte} & \SI{64}{\kilo\byte} & N/A & IS MPU9250 \\
Myo Armband          & N/A & N/A & N/A & IS MPU9150 \\
Apple Watch S7    & \SI{1.8}{\giga\hertz}& \SI{1}{\giga\byte} & Watch OS 9 & Unknown \\
Huawei Watch GT2 & \SI{200}{\mega\hertz} & \SI{32}{\mega\byte} & Lite OS 11 & STM LSM6DSO \\
\hline
\textbf{Name} & \textbf{{Start Freq}} & \textbf{{ADC Samples}} & \textbf{{Chirp Loops}} & \textbf{Idle Time}\\
\hline  TI IWR1843
             & \SI{77}{\giga\hertz}                 & 256           & 255                 & \SI{100}{\micro\second}
 \\
\hline
\end{tabular}
\begin{tablenotes}
    \item[*]IS: TDK InvenSense, STM: STMicroelectronics.
\end{tablenotes}
\end{threeparttable}
\label{tab:device}
\vspace{-0.1in}
\end{table}

\begin{figure}
    \centering
    \includegraphics[width=0.4\textwidth]{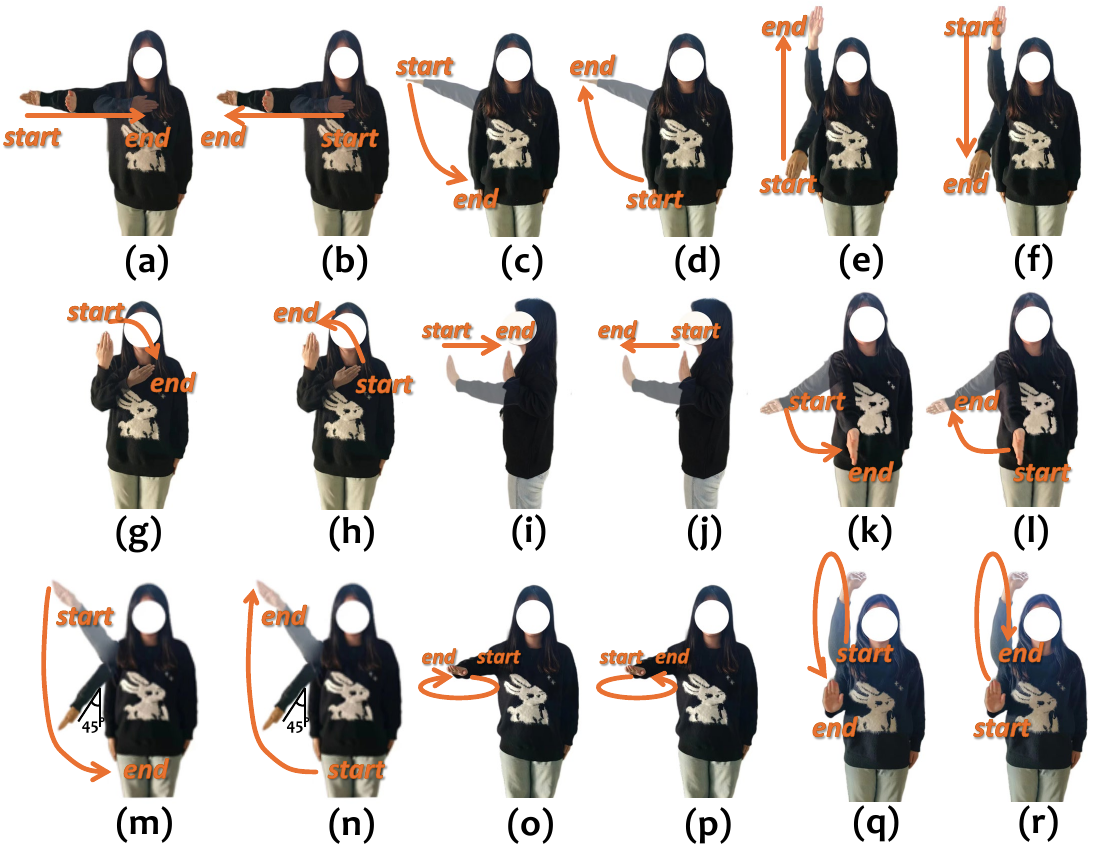}
    \caption{\textbf{Gestures in the dataset.} 18 distinct gestures, including lateral-to-front raises, lateral raises, front raises, forearm supination/pronation, push, pull, swipes, \SI{45}{\degree} lateral raises, horizontal rotations, and vertical rotations.}
    \label{fig:dataset}
    \vspace{-0.3in}
\end{figure}

\noindent\textbf{Data collection.} 
To validate \SystemName, we enlisted 30 volunteers consisting of 17 females and 13 males, ranging in age from 15 to 64 years. All participants were in good health and took part in a series of controlled experiments~\footnote{Ethical approval has been obtained from the corresponding organization.}. The data collection spanned over a three-month period.
Our study involved eighteen distinct gestures, as depicted in Fig.~\ref{fig:dataset}, encompassing a variety of wrist, elbow, and shoulder movements.
To test the I2R diffusion model, we randomly selected half of the participants (15 individuals). Each participant executed all eighteen gestures 15 times while positioned in front of the mmWave radar equipped with an IMU sensor. These sessions were conducted under diverse conditions: indoors, outdoors, and through-obstacle scenarios, as illustrated in Fig.~\ref{fig:dev}(d)(e)(f).
For the evaluation of gesture recognition, the remaining 15 participants were instructed to complete two distinct sets of gesture trials. The first set aimed at gathering data for the translation recognition model, required participants to perform each of the eighteen gestures 15 times while carrying mobile devices across different settings, including indoor, outdoor, and through-obstacle environments. The second set focused on collecting gesture recognition data, with participants repeating each gesture 15 times in the presence of the radar.

\vspace{-0.1in}

\subsection{Overall Performance}
\begin{figure}
\centering
\subfigure[Overall accuracy.]{
\begin{minipage}[t]{0.47\linewidth}
\centering
\includegraphics[width=0.97\linewidth]{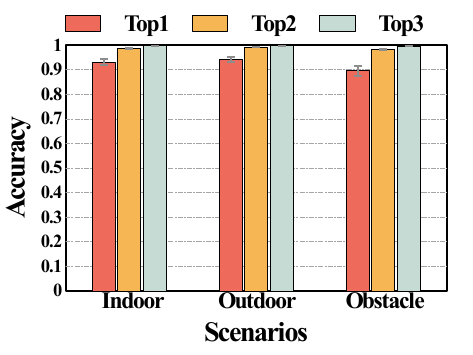}
\label{fig:eva1}
\vspace{-4mm}
\end{minipage}%
}
\hspace{-0.1in}
\subfigure[Comparison with baselines.]{
\begin{minipage}[t]{0.47\linewidth}
\centering
\includegraphics[width=0.97\linewidth]{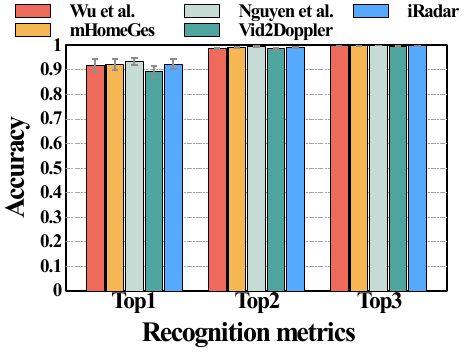}
\label{fig:base}
\vspace{-4mm}
\end{minipage}%
}
\vspace{-2mm}
\centering
\caption{\textbf{Overall Performance.}}
\vspace{-0.3in}
\end{figure}

\noindent\textbf{Overall accuracy.} 
Fig.~\ref{fig:eva1} provides the accuracy percentages for gesture recognition across indoor, outdoor, and through-obstacle scenarios, utilizing the recognition model trained with \SystemName. The top-N accuracy criterion indicates the rate at which the correct gesture is identified within the top-N selections.
Specifically, the Top-1 accuracy achieved in indoor settings was 93.1\%, while outdoor settings saw a slightly higher of 94.3\%. For through-obstacle conditions, the Top-1 accuracy is lower at 89.6\%, attributable to the attenuating effects of obstacles on mmWave signal propagation. For Top-2 accuracy, the values remained notably high at 98.7\% for indoor, 99.3\% for outdoor, and 98.3\% for through-obstacle scenarios. For Top-3 accuracies, with 99.8\% for indoor, 99.9\% for outdoor, and 99.7\% for through-obstacle conditions.
The results exhibit a high effectiveness of \SystemName across a diverse array of environmental conditions. 

\noindent\textbf{{Comparison with baselines.}} 
\huanqi{We compare \SystemName with three types of state-of-the-art gesture recognition systems.
(i) mmWave-based: Wu et al.'s system~\cite{wu2022toward}, utilizing mmWave Doppler heatmaps and mHomeGes~\cite{liu2020real}, based on mmWave point clouds; and (ii) IMU-based: Nguyen et al.'s IMU-based system~\cite{nguyen2021gesture}; (iii) video to mmWave translation: Vid2Doppler~\cite{ahuja2021vid2doppler} with video translated mmWave heatmaps. To ensure fairness, we adjusted each system to work best with our dataset. As presented in Fig.~\ref{fig:base}, \SystemName is only 1.2\% less accurate than the system by Nguyen et al., yet it is 0.7\% more accurate than the system by Wu et al., 0.2\% more accurate than mHomeGes, and 4\% more accurate than Vid2Doppler. The IMU-based system shows higher accuracy because wearables are attached to the body, resulting in lower environmental noise. The results show that \SystemName achieves comparable accuracy to the state-of-the-art systems.}

\begin{figure*}
\centering
\subfigure[Performance of I2R model.]{
\begin{minipage}[t]{0.24\linewidth}
\centering
\includegraphics[width=0.95\linewidth]{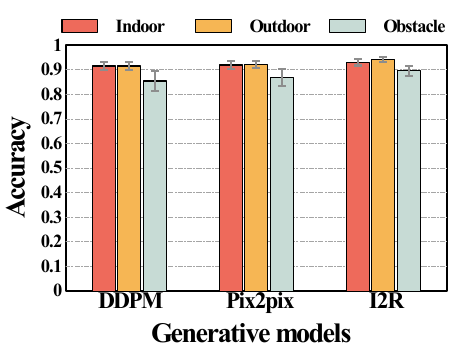}
\label{fig:eva5}
\vspace{-4mm}
\end{minipage}%
}%
\subfigure[Impact of doppler transformer.]{
\begin{minipage}[t]{0.24\linewidth}
\centering
\includegraphics[width=0.95\linewidth]{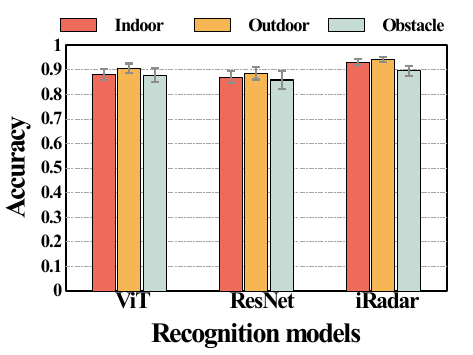}
\label{fig:eva6}
\vspace{-4mm}
\end{minipage}%
}%
\subfigure[Impact of MC-MWHE.]{
\begin{minipage}[t]{0.24\linewidth}
\centering
\includegraphics[width=0.95\linewidth]{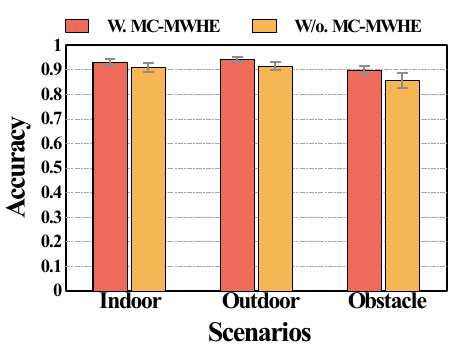}
\label{fig:eva7}
\vspace{-4mm}
\end{minipage}%
}%
\subfigure[Generative performance.]{
\begin{minipage}[t]{0.24\linewidth}
\centering
\includegraphics[width=0.95\linewidth]{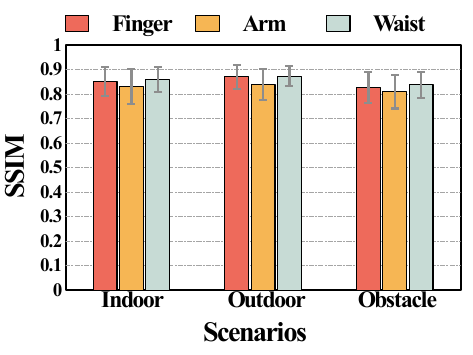}
\label{fig:eva2}
\vspace{-4mm}
\end{minipage}%
}%
\\
\vspace{-0.1in}
\subfigure[Cross-device generalization.]{
\begin{minipage}[t]{0.24\linewidth}
\centering
\includegraphics[width=0.95\linewidth]{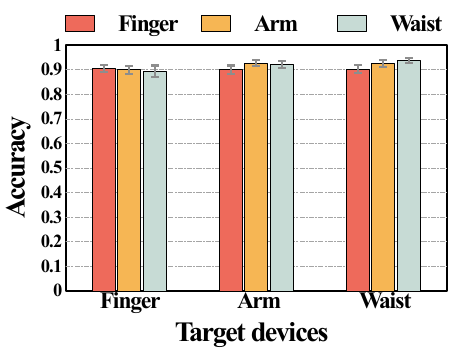}
\label{fig:eva1_gen}
\vspace{-4mm}
\end{minipage}%
}%
\subfigure[Impact of position.]{
\begin{minipage}[t]{0.24\linewidth}
\centering
\includegraphics[width=0.95\linewidth]{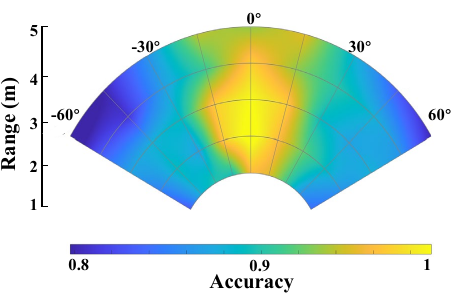}
\label{fig:eva3}
\vspace{-4mm}
\end{minipage}%
}%
\subfigure[Impact of user number.]{
\begin{minipage}[t]{0.24\linewidth}
\centering
\includegraphics[width=0.95\linewidth]{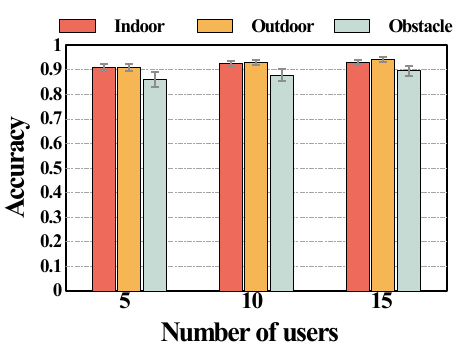}
\label{fig:eva4}
\vspace{-4mm}
\end{minipage}%
}%
\subfigure[Impact of sampling rate.]{
\begin{minipage}[t]{0.24\linewidth}
\centering
\includegraphics[width=0.95\linewidth]{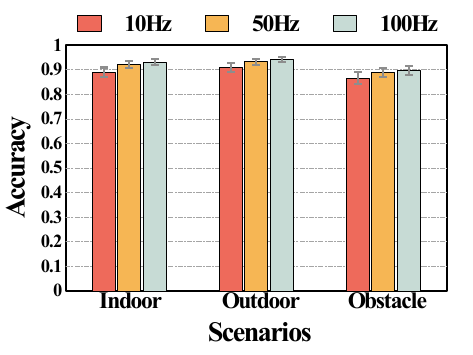}
\label{fig:eva8}
\vspace{-4mm}
\end{minipage}%
}%
\vspace{-2mm}
\centering
\caption{\textbf{Experimental results.}}
\vspace{-0.3in}
\end{figure*}

\subsection{Micro-Benchmark Evaluation}
\label{subsec:bench}

\noindent\textbf{Performance of I2R diffusion model.} 
The effectiveness of the I2R diffusion model is shown in Fig.~\ref{fig:eva5}, where we compare the accuracy utilizing the I2R model against those employing DDPM~\cite{ho2020denoising} and pix2pix~\cite{isola2017image}. For integrating the IMU inputs, both DDPM and pix2pix are accompanied by an inertial fusion module. The results indicate that I2R surpasses the comparative models in every environment. Notably, I2R achieves an increase in accuracy over DDPM by 1.64\%, 2.94\%, and 4.91\% in indoor, outdoor, and through-wall settings, respectively. When measured against pix2pix, we observe accuracy improvements of 1.20\%, 2.17\%, and 3.41\%.

\noindent\textbf{Evaluation on Doppler transformer.} 
We then assess the proposed gesture recognition model, the Doppler transformer, by benchmarking its accuracy against that of the Vision Transformer (ViT)\cite{dosovitskiy2020image} and the Residual Network (ResNet)\cite{he2016deep} across various settings, including indoor, outdoor, and through-wall scenarios. Fig.~\ref{fig:eva6} presents a comparison of the performance of different models in these environments.
The results show that \SystemName exceeds the mean accuracy of ViT and ResNet by margins of 4.13\% and 5.73\%, respectively. This substantial improvement is indicative of \SystemName's advanced capability to capture and interpret gesture-related features.


\noindent\textbf{Evaluation on MC-MWHE method.}
The impact of our proposed mmWave heatmap enhancement technique, MC-MWHE, was assessed by comparing performance metrics with and without the application of this method. As depicted in Fig.~\ref{fig:eva7}, the deployment of MC-MWHE resulted in accuracy improvements of 2.3\%, 3.3\%, and 4.7\% for indoor, outdoor, and through-obstacle scenarios, respectively. These enhancements underscore the effectiveness of MC-MWHE in noise reduction and the overall refinement of the recognition.

\noindent\textbf{\huanqi{Generative information loss.}} 
The generative results of the I2R diffusion model, which transforms IMU spectrograms into mmWave radar heatmaps, are presented in Fig.~\ref{fig:eva2}. We use the Structural Similarity Index Measure (SSIM)~\cite{wang2004image} to assess the similarity between the generated heatmaps and their authentic counterparts. The I2R model demonstrates superior performance, with SSIM scores indicating high levels of similarity in different scenarios: in indoor settings, the model achieves 85.21\%, 83.27\%, and 86.01\% for three different wearable devices; in outdoor environments, SSIM scores are 87.12\%, 84.07\%, and 87.39\%; and through-obstacle conditions resulted in 82.92\%, 81.18\%, and 83.98\%.
These results are indicative of the model's ability to produce outputs that are closely aligned with the original mmWave signal.

\noindent\textbf{Generalization ability.}  
The ability of \SystemName to adapt to diverse environments and wearable device placements is rigorously evaluated. Tab.~\ref{tbl:type2} displays the system's recognition capabilities across a range of IMU and mmWave data collection scenarios. The table illustrates that \SystemName has a consistently high level of accuracy in various settings, with all Top-3 accuracy surpassing 99.5\%. It is observed, however, that performance slightly dips in through-obstacle scenarios due to the impact of obstructions on the mmWave signal. Conversely, the system excels in outdoor scenarios, benefiting from the absence of obstructions and reduced interference.
To further assess \SystemName's adaptability to different wearable device positions, we establish a baseline using a model from one wearable position and then enhance models for two additional positions with ten epochs of further training. The results, as depicted in Fig.~\ref{fig:eva1_gen}, show that each position attains an accuracy of over 89.5\%. This underscores the model's robust generalization across wearable positions.

\begin{table}[t]
\caption{\textbf{Cross-scenario results.}}
\fontsize{6.3pt}{7.2pt}\selectfont 
\begin{threeparttable}
\setlength{\tabcolsep}{4pt}
\begin{tabular}{ccc|ccc|ccc}
\toprule

\multicolumn{3}{c|}{\textbf{IMU Sce.\tnote{*}}} & \multicolumn{3}{c|}{\textbf{mmWave Sce.}} &  \multicolumn{3}{c}{\textbf{Accuracy}} \\

\cmidrule(rl){1-3}
\cmidrule(rl){4-6}
\cmidrule(rl){7-9}

\multicolumn{1}{c}{\rotatebox{0}{I.D.}}& \multicolumn{1}{c}{\rotatebox{0}{O.D.}}& \multicolumn{1}{c|}{\rotatebox{0}{T.O.}}  
& \multicolumn{1}{c}{\rotatebox{0}{I.D.}}& \multicolumn{1}{c}{\rotatebox{0}{O.D.}}& \multicolumn{1}{c|}{\rotatebox{0}{T.O.}} 
& \multicolumn{1}{c}{\rotatebox{0}{Top-1}}& \multicolumn{1}{c}{\rotatebox{0}{Top-2}}& \multicolumn{1}{c}{\rotatebox{0}{Top-3}}

\\
\midrule
\CIRCLE& \Circle& \Circle      & \CIRCLE& \Circle& \Circle   & \timebar{100}{93.12}\% & \timebar{100}{98.54}\% & \timebar{100}{99.71}\%                                                         \\
\Circle& \CIRCLE & \Circle                              & \CIRCLE& \Circle& \Circle    & \timebar{100}{93.21}\% & \timebar{100}{98.67}\% & \timebar{100}{99.81}\%    
   \\
\Circle& \Circle& \CIRCLE                               & \CIRCLE& \Circle& \Circle    & \timebar{100}{89.32}\% & \timebar{100}{98.22}\% & \timebar{100}{99.63}\%      
\\
\CIRCLE& \Circle & \Circle                        & \Circle& \CIRCLE& \Circle        & \timebar{100}{92.43}\% & \timebar{100}{98.99}\% & \timebar{100}{99.91}\%                                                            
\\
\Circle& \CIRCLE & \Circle                              & \Circle& \CIRCLE& \Circle    & \timebar{100}{94.32}\% & \timebar{100}{98.93}\% & \timebar{100}{100}\%    
\\
\Circle& \Circle & \CIRCLE                               & \Circle& \CIRCLE& \Circle    & \timebar{100}{88.23}\% & \timebar{100}{98.31}\% & \timebar{100}{99.55}\%    
\\
\CIRCLE& \Circle & \Circle                          & \Circle& \Circle& \CIRCLE      & \timebar{100}{89.47}\% & \timebar{100}{98.35}\% & \timebar{100}{99.67}\%                                                            \\
\Circle& \CIRCLE & \Circle                          & \Circle& \Circle& \CIRCLE      & \timebar{100}{88.26}\% & \timebar{100}{98.23}\% & \timebar{100}{99.54}\%      
\\
\Circle& \Circle& \CIRCLE                           & \Circle& \Circle& \CIRCLE      & \timebar{100}{89.71}\% & \timebar{100}{98.41}\% & \timebar{100}{99.81}\%  
\\

\bottomrule
\end{tabular}
\begin{tablenotes}
    \item[*]\CIRCLE \thinspace for chosen, \Circle \thinspace for unchosen, I.D.:indoor, O.D.: outdoor, T.O.: through-obstacle.
\end{tablenotes}
\end{threeparttable}

\label{tbl:type2}
\vspace{-0.3in}
\end{table}

\noindent\textbf{Evaluation on user position.} 
Fig.~\ref{fig:eva3} illustrates the influence of the user's relative position to mmWave radar on the accuracy of \SystemName, with distances ranging from \SIrange{1}{5}{\meter} and angular displacement spanning \SIrange{-60}{60}{\degree}. It is observed that the system's accuracy initially increases with distance but subsequently diminishes. To be precise, average accuracy increases by 12.43\% when the distance extends from \SI{1}{\meter} to \SI{3}{\meter} and then declines by 7.56\% as the distance further grows to \SI{5}{\meter}. This suggests that at closer proximities, the mmWave radar's sensitivity to user orientation can negatively impact recognition accuracy. Additionally, the system exhibits better performance within \SIrange{-30}{30}{\degree}, because users are more prominently within the radar's optimal sensing radius.

\noindent\textbf{\huanqi{Evaluation on user number.}} 
In our investigation of the effect of user group size on recognition accuracy within \SystemName, we observe notable trends as illustrated in Fig.~\ref{fig:eva4}. Accuracy exhibits a positive correlation with group size, where an increase from 5 to 10 members resulted in a 2.45\% improvement in accuracy, and a further increase to 15 members led to an additional 1.13\% improvement. This enhancement can be attributed to the greater information of distinguishable gestures present within larger groups, which contributes to the system's ability to correctly recognize them. 
\SystemName achieves over 88\% accuracy even in groups as small as five.


\noindent\textbf{Evaluation on sampling rate.} 
We then delve into the relationship between the sampling rate and the accuracy of recognition results in \SystemName. We present our findings for sampling rates of \SI{10}{\hertz}, \SI{50}{\hertz}, and \SI{100}{\hertz}. Fig.~\ref{fig:eva8} illustrates that when the sampling rate is reduced from \SI{100}{\hertz} to \SI{50}{\hertz}, there is a marginal decrease in accuracy by approximately 1.13\%. A further reduction of the sampling rate to \SI{10}{\hertz} leads to a more pronounced decline of about 2.60\% in accuracy. This significant drop is largely due to the omission of vital gesture data that occurs at lower sampling rates, thereby negatively affecting the system's ability to recognize patterns accurately. 

\begin{figure}
    \centering
    \includegraphics[width=0.46\textwidth]{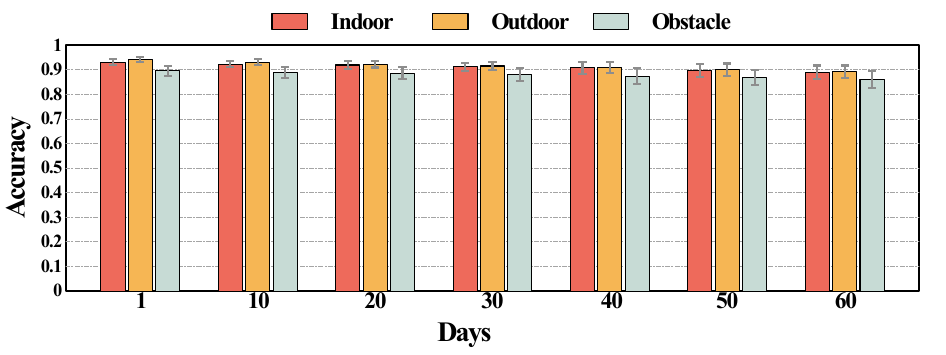}
    \vspace{-1mm}
    \caption{\textbf{Long-term performance.}}
    \vspace{-0.3in}
    \label{fig:longterm}
\end{figure}

\noindent\textbf{Evaluation on time-varying performance.} 
Human gestures are inherently variable, which requires a thorough evaluation of the temporal reliability of \SystemName. We engaged five individuals using a smartwatch continuously over two months, with evaluations every ten days. As shown in Fig.~\ref{fig:longterm}, a gradual decrease in recognition accuracy was observed. More precisely, there was an average reduction in accuracy of 2.17\% at the one-month mark and an additional 2.55\% reduction by the end of the second month.
The observed decline can be ascribed to natural variations in the users' physiological characteristics, such as muscle tone and joint flexibility, which subtly alter gesture patterns. Despite these changes, \SystemName maintained an average accuracy rate of 88.07\% after 60 days, showcasing its resilience. To maintain efficacy over time, methods like continuous learning are recommended~\cite{anderson2008theory}.

\section{Related work}


\noindent\textbf{Cross-modality translation for mmWave.}
Current research in cross-modality translation for mmWave predominantly focuses on generating mmWave signals from videos~\cite{ahuja2021vid2doppler,deng2023midas,zhang2022synthesized}. Notably, Vid2Doppler~\cite{ahuja2021vid2doppler} converts human activity captured in videos into highly realistic synthesized mmWave radar data via a transformer-based model. Similarly, the Midas system~\cite{deng2023midas} employs an enhanced transformer model in conjunction with VS-Net to generate believable radar data and pinpoint salient video segments. SynMotion~\cite{zhang2022synthesized} uses existing video datasets to translate video information into synthetic mmWave data for human motion sensing.
Despite their advancements, these video-based methods are vulnerable to environmental factors such as occlusions and varying lighting conditions. To overcome these challenges, \SystemName leverages less susceptible IMU data for modality translation.

\noindent\textbf{mmWave-based gesture recognition.}
Current mmWave-based gesture recognition methods are broadly divided into heatmap-based approaches~\cite{yan2023mmgesture,yuan2022real} and point cloud-based approaches~\cite{palipana2021pantomime, liu2020real,salami2022tesla}. Yuan et al.~\cite{yuan2022real} utilized CNNs to recognize digital gestures from hand motion trajectories depicted in heatmaps. On the other hand, point cloud-based methods, such as Pantomime~\cite{palipana2021pantomime}, process sparse 3D point clouds derived from radar signals through deep learning frameworks. Similarly, mHomeGes~\cite{liu2020real} recognizes arm gestures in real-time by reconstructing point clouds and employing shallow neural networks for classification.
Despite the progress, point cloud-based methods often struggle with sparsely filled point clouds that fail to capture gestures with high fidelity, as evaluated by recent study~\cite{jiang2023evaluating}. Additionally, traditional heatmap-based approaches that rely on CNNs may be constrained by their local receptive fields and shared weights, which limits their capacity to understand gestural data.

\section{Conclusion}
\label{sec:conclusion}
In this work, we introduce \SystemName, which, to the best of our knowledge, is the first framework for cross-modal IMU-to-mmWave gesture recognition that circumvents the necessity of additional mmWave hardware installation and obviates the need for explicit data collection, substantially alleviating the service providers' burden. \SystemName encompasses a diffusion-driven translation method, a novel mmWave heatmap enhancement technique,  and a Doppler transformer recognition algorithm. Our comprehensive evaluation reveals that \SystemName achieves an average Top-3 accuracy rate of 99.82\%.

\section*{Acknowledgment}
This project was supported by National Key R\&D Program of China (Grant No. 2023YFE0208800), the Research Grants Council of the Hong Kong SAR, China (Project No. CityU 11202124 and CityU 11201422), NSF of Guangdong Province (Project No. 2024A1515010192), the Innovation and Technology Commission of Hong Kong (Project No. MHP/072/23). *Weitao Xu is the corresponding author.

\bibliographystyle{IEEEtran}
\bibliography{reference}

\end{document}